\documentclass[sigconf, nonacm]{acmart}

\usepackage{svg}
\usepackage{booktabs}
\usepackage{multirow}
\usepackage{textcomp}
\usepackage{hyperref}
\usepackage{balance}
\hypersetup{
    colorlinks=true,
    linktocpage=true,
    linkcolor=blue,
    filecolor=magenta,      
    urlcolor=cyan,
}
\AtBeginDocument{%
  \providecommand\BibTeX{{%
    \normalfont B\kern-0.5em{\scshape i\kern-0.25em b}\kern-0.8em\TeX}}}

\setcopyright{acmcopyright}
\copyrightyear{2021}
\acmYear{2021}
\acmConference[WSDM WebTour'21]{ACM WSDM Workshop on Web Tourism (WSDM Webtour'21}{March 12, 2021}{Jerusalem, Israel}
\acmBooktitle{ACM WSDM Workshop on Web Tourism (WSDM Webtour'21), March 12, 2021, Jerusalem, Israel}
\acmPrice{}
\acmDOI{}
\acmISBN{}
\begin{document}

\title{Data Augmentation Using Many-To-Many RNNs \\  for Session-Aware Recommender Systems}

\author{Martín Baigorria Alonso}
\email{martinbaigorria@gmail.com}
\authornote{Paper submitted as an independent researcher.}

\begin{abstract}

The ACM WSDM WebTour 2021 Challenge organized by Booking.com focuses on applying Session-Aware recommender systems in the travel domain. Given a sequence of travel bookings in a user trip, we look to recommend the user's next destination. To handle the large dimensionality of the output's space, we propose a many-to-many RNN model, predicting the next destination chosen by the user at every sequence step as opposed to only the final one. We show how this is a computationally efficient alternative to doing data augmentation in a many-to-one RNN, where we consider every subsequence of a session starting from the first element. Our solution achieved 4th place in the final leaderboard, with an accuracy@4 of 0.5566.
\end{abstract}



\begin{teaserfigure}
  \includegraphics[width=\textwidth]{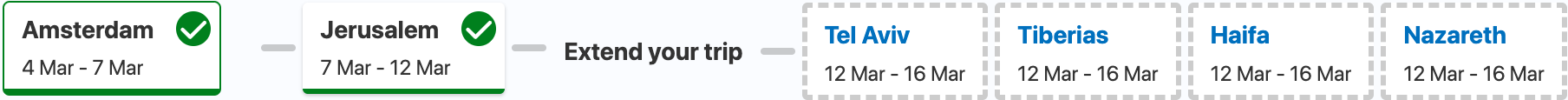}
  \caption{Multi-Destinations Trip Recommender on Booking.com}
  \label{fig:booking-recommender}
\end{teaserfigure}
  
\maketitle

\section{Introduction}

Session-Aware recommender systems model the sequential decision process of a user in the context of a session, also considering past user actions or attributes \cite{DBLP:journals/corr/abs-1802-08452, DBLP:journals/corr/abs-1902-04864}. These systems have been essential for the growth of many e-commerce and content companies that need to organize a vast catalog of options into a relevant and easily manageable subset by the user \cite{10.1109/MIC.2003.1167344, Resnick94grouplens:an, 10.1145/2959100.2959190, 10.1145/2843948, 10.1145/3240323.3240354}.


The ACM WSDM WebTour 2021 Challenge organized by Booking.com \cite{booking2021challenge} proposes a Session-Aware recommender systems problem that focuses on using a dataset of booking sequences and contextual information to make the best possible recommendation for a user in real-time. The quality of the recommendations is measured using recall@4. Booking.com currently has a recommender system for this problem in production (e.g., Figure \ref{fig:booking-recommender}), which enforces once again the importance of the problem.

Hidasi et al. \cite{hidasi2016sessionbased} has recently proposed the use of Recurrent Neural Network (RNN) based models to overcome some of the difficulties other factor models or neighborhood-based approaches can have when modeling sparse sequential data. These models have been successful in other domains such as speech recognition \cite{6296526} and natural language processing \cite{mikolov2010recurrent, sutskever2014sequence, DBLP:journals/corr/WuSCLNMKCGMKSJL16}.

This manuscript proposes an approach to handling a high dimensional output space in RNN based Session-Aware recommender systems. Our main contributions to this problem include extending the many-to-one configuration typically-used RNN-based recommender systems to the many-to-many configuration. Instead of predicting the next booking by a user given a sequence of bookings, we predict the next booking at every time step. We show how this is a computationally efficient alternative to doing data augmentation in a many-to-one RNN, where we consider every subsequence of a session starting from the first element. Our empirical results show how this model can outperform the many-to-one RNNs by a significant margin. We also show how this extension biases the learning problem towards shorter trips and propose a correction by weighting the model's loss function.

We organize the rest of the paper as follows: In Section \ref{section:overview}, we first give an overview of the challenge, the dataset provided by Booking.com, and the evaluation metric to be optimized. We then describe the many-to-many RNN architecture we used. We focus on explaining our design choices' motivations and the approach we followed for training and inference (in Section \ref{section:approach}). Lastly, in Section \ref{section:experimental-setup} we discuss both the implementation details we used to iterate quickly and our experiments' results. Our code and documentation are available at \href{https://github.com/mbaigorria/booking-challenge-2021-recsys}{https://github.com/mbaigorria/booking-challenge-2021-recsys}.

\section{Challenge Overview}
\label{section:overview}

The ACM WSDM WebTour 2021 challenge proposed by Booking.com is an instance of the multi-destinations trip planning problem. Using an anonymized dataset of bookings, the goal is to recommend the next possible destination to a user in the context of a session.

\subsection{Data}

The dataset is composed of over a million anonymized reservations. Each hotel reservation is a part of a user's trip (identified by utrip\_id) and includes multiple features such as the city, country, check-in, and checkout times. In the test set, these trips include at least four consecutive reservations. An overview of the dataset statistics is shown in Table \ref{table:descriptive}.

\vspace{-0.1cm}
\begin{table}[H]
\resizebox{\columnwidth}{!}{
\begin{tabular}{llcccrrr}
\hline
\multicolumn{1}{c}{\multirow{2}{*}{Type}} &
  \multicolumn{1}{c}{\multirow{2}{*}{Users}} &
  \multirow{2}{*}{Sessions} &
  \multirow{2}{*}{Cities} &
  \multirow{2}{*}{Accomodations} &
  \multicolumn{3}{c}{Session length} \\ \cline{6-8} 
\multicolumn{1}{c}{} & \multicolumn{1}{c}{} &                             &                            &                               & Min & Max & Median \\ \hline
Train                & 200,153              & \multicolumn{1}{r}{217,686} & \multicolumn{1}{r}{39,901} & \multicolumn{1}{r}{1,166,835} & 1   & 48  & 5      \\
Test                 & 68,502               & \multicolumn{1}{r}{70,662}  & \multicolumn{1}{r}{21,305} & \multicolumn{1}{r}{378,667}   & 4   & 44  & 5      \\ \hline
\end{tabular}
}
\caption{Descriptive statistics of the training and test sets}
\label{table:descriptive}
\end{table}
\vspace{-0.8cm}

For each session in the test set, the city and country of the last booking are concealed. However, at inference time, other variables such as the check-in day of the next booking are also available.

\subsection{Evaluation}

For every trip in the test set, the correct city to be recommended is concealed. The metric used to evaluate the recommendations was precision@4. We can understand this metric as the percentage of times the correct city was in the top 4 recommendations for each trip. For this problem, the metric is equivalent to the recall@4, since there is only one relevant item to recommend per trip.




\section{Approach \& Motivation}
\label{section:approach}

We designed our approach to be sequence and distance aware. Predicting the next destinations is a problem that must capture the notion of physical distance between cities and the sequential nature of the user's decision-making process. We should also be able to handle variable-length user trips. It is also worth remembering user trips may be censored, as users do not necessarily book every trip on the platform. 

Given some highly dimensional categorical variables such as affiliate\_id or city\_id, we should also learn latent representations (also known as embeddings) to explicitly avoid using one-hot encodings. The large dimensionality of the output space can also be challenging. In this specific case, there are 39,901 possible cities to recommend. In the training set, only 52\% of these cities appear as the last element of a trip. If we are not careful, any softmax-based model to predict the sequence's final booking is unlikely to make all cities predictable.

In Section \ref{subsection:architecture}, we explain how we tackled these challenges, the problems and trade-offs we faced, and some interesting future research directions.

\vspace{-0.4cm}
\begin{figure}[H]
  \centering
  \includegraphics[width=1\columnwidth]{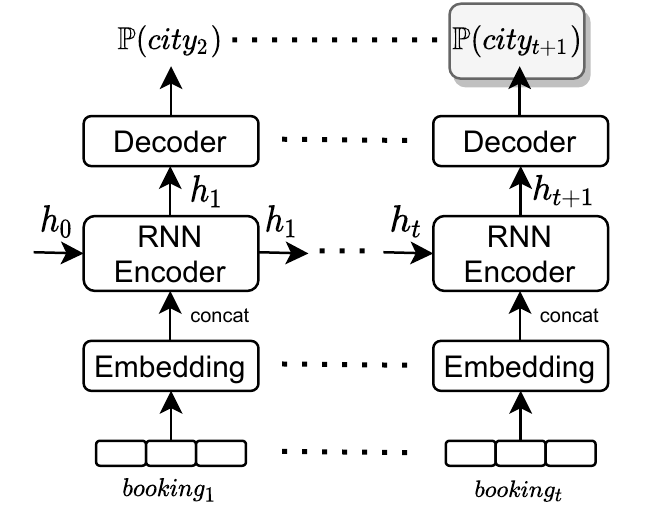}
  \caption{Many-to-many RNN based model architecture}
  \label{fig:architecture}
\end{figure}
\vspace{-0.4cm}

\subsection{Architecture}
\label{subsection:architecture}
Our neural network architecture can be divided into three components. First, our model concatenates the embeddings of all features for each user trip booking. These concatenated vectors are then fed into a many-to-many RNN encoder, outputting the next city's probability mass function given the previous cities at every time step. We finally recommend the top 4 cities in the probability mass function of the final step. You can see our general architecture diagram in Figure \ref{fig:architecture}.


\subsubsection{\textbf{Feature engineering and representation}}

\subsubsection*{Feature engineering}

Using the provided dataset, we created 14 features by using the available information of both the current and the next bookings. We were cautious to not introduce any type of data leakage. The variables used were:
\begin{itemize}
\item City and country of the current booking.
\item Country in which the current booking  made.
\item Country in which the next booking is made.
\item Check-in day, month, and year of the current booking.
\item Check-in day of the next booking.
\item Duration of stay (days) of the current booking.
\item Duration of stay (days) of the next booking.
\item Device class of the current booking.
\item Transition days between a checkout and the next check-in.
\item Current and next booking affiliate\_id (e.g. direct, third-party referral, paid search engine, etc.).
\end{itemize}

\noindent The transition days between bookings can potentially help us capture censored bookings. The model can learn that an extended transition time could mean the user booked on another platform, depending on the city to be recommended.

\subsubsection*{Feature representation}

We embed all categorical and numerical features. Numerical features have a low cardinality, so they should be easily learned by the model. This allowed us not to have to worry about feature scaling. The embedding dimension of all numerical features is set to 10 and for categorical features to 25 except for the device\_id (5) and the city\_id (128). To input these features into our model, we then concatenate all the embeddings at each time step. We also tried to merge all feature embeddings through multiplication, but our model degraded. This is similar to the findings by Mizrahi et al. for a similar dataset \cite{Mizrachi2019CombiningCF}.

\subsubsection{\textbf{Encoder}}

We use a 2 layered RNN encoder to represent the context of the user at every sequence time step. In a nutshell, an RNN can process a variable length sequence as follows:

\vspace{-0.1cm}
\begin{equation}
h_t = g(h_{t-1}, f_t, \theta)
\end{equation}

\noindent where g is a recurrent transition function parameterized by $\theta$ (e.g. GRU, LSTM), $h_t$ is the hidden state at time $t$ and $f_t$ is the input vector at time $t$ (the concatenated booking embeddings in our case). From the hidden state at time $t$, we can not only get the hidden state at $t+1$, but can also use a decoder to generate a probability mass function over cities in $t+1$. We experiment with both GRU and LSTM encoders \cite{10.1162/neco.1997.9.8.1735, DBLP:journals/corr/ChoMGBSB14}. As others have previously pointed out, we found that GRUs tend to outperform LSTMs in some recommendation settings \cite{hidasi2016sessionbased, Mizrachi2019CombiningCF}.

Instead of using a many-to-one RNN, we decided to use a many-to-many architecture to predict the next booked city at each time step. This many-to-many architecture has the same number of parameters as the many-to-one architecture. We apply a decoder over all the hidden states for each sequence instead of just the last one. This is identical to training a many-to-one architecture with an augmented dataset containing all session subsequences beginning from the first element in a single batch. However, for a session of length $n$, the many-to-one architecture must do $n$ forward passes with a number of sequential operations in $\mathcal{O}(n^2)$ to get all the hidden states. On the other hand, the many-to-many model computes the same hidden states but linearly in $n$.

We apply recurrent dropout with a probability of 0.1 to both of the recurrent encoders. We also use dropout on the concatenated input embeddings with a probability of 0.3. This is done not only to regularize our model \cite{srivastava2014dropout} but also to model the censorship in the data by partially omitting some features in the different time steps.

\subsubsection{\textbf{Decoder}}

The decoder maps the hidden state $h_t$ of the RNN encoder every time step $t$ to a probability mass function over the 39,901 cities. We have experimented with the following two parameterizations of the decoder:

\begin{enumerate}
\item \textbf{Feedforward Neural Network}: This layer is parameterized with an input size equal to the number of hidden units in the RNN encoder and an output size equal to the number of cities. We then apply a softmax to the final output at each time step to get a probability mass function over cities.
\item \textbf{Tied weights between city embeddings and output \\ layer}: We set the dimension of the encoder hidden state $h_t$ to be equal to the dimension of the city embeddings. We can then understand the encoder as learning a vector in a similar manifold as the cities.  We then calculate the product between the hidden states and the transposed embedding matrix of cities, applying a softmax after to get a probability mass function over all cities at every time step. This is equivalent to parameterizing the feedforward encoder mentioned above with the weights of the city embedding matrix. Some implementations of Word2Vec \cite{DBLP:journals/corr/MikolovSCCD13} use a similar trick, where the input word and context word embeddings are parameterized with the same weights. It has recently been shown that this can also be interpreted as a form of data augmentation \cite{DBLP:journals/corr/InanKS16}.
\end{enumerate}

We concluded that tying the city embeddings with the output layer does not only lead to a faster training time because the network has a lower number of parameters, but it also has a better performance across different encoder configurations. You can see Table \ref{table:experiments} for more details.

\subsection{Loss function}

We calculate the cross-entropy loss of the probability mass function over cities at every sequence step for each trip. Since we train the model in batches, we then average all these cross-entropy losses for every trip in a batch.



We have validated that the longer a user trip is, the easier it is for our model to predict the next booking. Table \ref{table:sequence_distribution} shows the sequence length distribution for different datasets once we remove the last observation in every sequence. We do this since we do not have information about the next booking. The last two datasets concatenate both the training and the test set. To clarify, concatenating both the training and the test sets provided in the challenge is possible since the actual leaderboard targets are concealed.

\begin{table}[H]
\begin{tabular}{@{}ccccc@{}}
\toprule
\begin{tabular}[c]{@{}c@{}}Session\\ length\end{tabular} &
  Train &
  Test &
  Train + Test &
  \begin{tabular}[c]{@{}c@{}}Train + Test\\ subsequences\end{tabular} \\ \midrule
1  & 0.001 & -     & 0.001 & 0.243 \\
2  & 0.003 & -     & 0.113 & 0.243 \\
3  & 0.452 & 0.453 & 0.398 & 0.215 \\
4  & 0.229 & 0.231 & 0.204 & 0.119 \\
5  & 0.126 & 0.127 & 0.114 & 0.069 \\
6  & 0.074 & 0.076 & 0.066 & 0.041 \\
7  & 0.044 & 0.042 & 0.040 & 0.025 \\
8  & 0.027 & 0.027 & 0.024 & 0.016 \\
9  & 0.016 & 0.016 & 0.015 & 0.010 \\
10 & 0.010 & 0.010 & 0.009 & 0.006 \\
> 10 & 0.028 & 0.027 & 0.026 & 0.006 \\ \midrule
\multicolumn{1}{l}{\begin{tabular}[c]{@{}l@{}}Total\\ sequences\end{tabular}} &
  217,573 &
  70,662 &
  288,235 &
  \textbf{\begin{tabular}[c]{@{}c@{}}1,186,491\\ (\textgreater{}4x more)\end{tabular}} \\ \bottomrule
\end{tabular}
\caption{Session length distribution for different types of datasets.}
\label{table:sequence_distribution}
\end{table}
\vspace{-0.6cm}

When considering all subsequences, the session length distribution becomes more concentrated on shorter sessions. This type of data augmentation comes at a cost since the underlying data generating process may be different for short and long trips.

Shorter sequences will also dominate the gradient estimates for each batch. We can correct this bias by weighting our loss function's cross-entropy outputs by the reverse cumulative frequency of the dataset sequence length. This way, each length's sequences would have the same contribution to gradient updates as in the non-augmented dataset. Our experiments in Table \ref{table:experiments} show weighting the loss function degraded the model's performance. However, this could be related to how we distributed the trips in a batch to train the model. This will be further discussed in Section \ref{section:results}.

It is also worth pointing out that augmenting the dataset explicitly has the risk of data leakage, using a validation trip that our model has already seen during training time. The many-to-many model avoids this by design, processing all subsequences of a session in one forward pass inside a single batch.


\subsection{Training and inference}

We trained a stratified ten-fold average cross-validation ensemble using the Adam optimizer \cite{kingma2015adam} with a learning rate of $10^{-3}$ and a batch size of 256. We trained each model for 50 epochs and used the recall@4 on the validation fold to pick the best model. 

We obtained some additional data by unifying the training and the test set. We were careful to split the dataset into folds with a similar sequence length distribution. We considered the last probability mass function at inference time and extracted the top 4 recommendations with the maximum probability, precisely as in a many-to-one model.

\section{Experimental Results \& Discussion}
\label{section:experimental-setup}

In this section, we dive into the implementation details and empirical evaluations of our proposed recommendation approach.

\subsection{Implementation}

Our model was implemented in PyTorch \cite{NEURIPS2019_9015}, using an NVIDIA\textsuperscript{\textregistered} Tesla\textsuperscript{\textregistered} V100 GPU for training. The following approaches allowed us to lower the training time per model to below 10 minutes. We trained with a large batch size of 256, sorting the sequences by length before batching to avoid zero padding. We did not explore the hyperparameters we used thoroughly, from the embedding sizes to the learning rate and batch size. We believe additional performance gains are possible by optimizing these hyperparameters.

We pre-loaded all batches in GPU, and when doing 10-fold cross-validation, we were careful to keep the folds as balanced as possible, also shuffling the batches at every epoch. Using a many-to-many RNN architecture instead of doing explicit data augmentation allowed us to process four times the amount of sequences at roughly the same computational cost.

\begin{table}[H]
\resizebox{\columnwidth}{!}{
\begin{tabular}{@{}cccrr@{}}
Model Type &
  \begin{tabular}[c]{@{}c@{}}Recurrent\\ Unit\end{tabular} &
  \multicolumn{1}{r}{\begin{tabular}[c]{@{}r@{}}Tie Encoder\\ \& Decoder\end{tabular}} &
  Weight Type &
  Accuracy@4 \\ \midrule
\multirow{10}{*}{Many To Many} &
  \multirow{6}{*}{GRU} &
  \multirow{4}{*}{True} &
  \multicolumn{1}{c}{UNWEIGHTED} &
  \textbf{0.5345*} \\
 &                       &                        & WEIGHTED   & 0.5313*       \\ \cmidrule(l){4-5} 
 &                       &                        & UNWEIGHTED & \underline{0.5202} \\
 &                       &                        & WEIGHTED   & 0.5196       \\ \cmidrule(l){3-5} 
 &                       & \multirow{2}{*}{False} & WEIGHTED   & 0.5119       \\
 &                       &                        & UNWEIGHTED & 0.5094       \\ \cmidrule(l){2-5} 
 & \multirow{4}{*}{LSTM} & \multirow{2}{*}{True}  & UNWEIGHTED & 0.5166       \\
 &                       &                        & WEIGHTED   & 0.5125       \\ \cmidrule(l){3-5} 
 &                       & \multirow{2}{*}{False} & UNWEIGHTED & 0.5161       \\
 &                       &                        & WEIGHTED   & 0.5128       \\ \midrule
\multirow{8}{*}{Many To One} &
  \multirow{4}{*}{GRU} &
  \multirow{2}{*}{True} &
  WEIGHTED &
  0.5064 \\
 &                       &                        & UNWEIGHTED & 0.5048       \\ \cmidrule(l){3-5} 
 &                       & \multirow{2}{*}{False} & WEIGHTED   & 0.5005       \\
 &                       &                        & UNWEIGHTED & 0.4996       \\ \cmidrule(l){2-5} 
 & \multirow{4}{*}{LSTM} & \multirow{2}{*}{True}  & WEIGHTED   & \underline{0.5078} \\
 &                       &                        & UNWEIGHTED & 0.5026       \\ \cmidrule(l){3-5} 
 &                       & \multirow{2}{*}{False} & UNWEIGHTED & 0.5071       \\
 &                       &                        & WEIGHTED   & 0.5017       \\ \bottomrule
\end{tabular}
}
\caption{Results for different configurations of our model. Models with an asterisk correspond to an average ten-fold cross-validation ensemble.}
\label{table:experiments}
\end{table}
\vspace{-0.6cm}

\subsection{Results}
\label{section:results}

Based on the implementation discussed in Section \ref{section:experimental-setup}, we measured the performance of our model's different configurations by building a new test set of 5,600 trips with 24,400 bookings from the training set. This test set had a similar trip length distribution as the challenge organizers' original test set.

Our results confirm GRUs seem to perform better than LSTMs for this task. We can also see the benefit of using many-to-many RNNs, which outperform many-to-one RNNs across all configurations. Tying the encoder and decoder weights also led to higher performance.

It is natural to wonder why the weighted version of the model did not perform as well as the unweighted version. We believe that this is related to our implementation. When we sorted the trips to avoid zero-padding, we biased each gradient update's mean and variance. Even if every batch has a fixed number of trips, this does not imply that it has the same number of augmented sequences. This can be solved using variable-sized batches, or avoiding sorting altogether.

\section{Conclusions}

One of the main challenges in Session-Aware recommender systems is effectively handling a large dimensional output space. This paper described how a many-to-many RNN could outperform the many-to-one configuration in such a setting. We showed how this extension is equivalent to doing data augmentation in a many-to-one RNN, with the pitfall of potentially biasing our gradient estimates. We finally showed the advantages of tying the feature embedding weights with the parameterization of the decoder.

The effect of having an uneven sequence length distribution in every batch and how to effectively handle the bias introduced by considering all subsequences in the many-to-many architecture can be an interesting future research direction. Attempting to do multi-task learning at every time step by also predicting other future attributes like the country might further boost our predictive performance. 

In the dataset provided by Booking.com, around 12\% of the users had multiple bookings. We attempted to model this by concatenating user trips with a separator token without success. This context can potentially be modeled using Hierarchical RNNs, or Transformer based architectures \cite{10.1145/3109859.3109896, NIPS2017_3f5ee243, sun2019bert4rec}.

\begin{acks}
We want to thank Sole Pera, Roberto Martín Pozzer, Leonardo Baldassini, Fabricio Previgliano, and Charlie Giudice for their useful comments on an earlier version of this work. We would also like to thank the ACM WSDM WebTour 2021 workshop organizers for the opportunity to participate in such an exciting challenge.
\end{acks}

\bibliographystyle{ACM-Reference-Format}
\balance
\bibliography{bibfile}

\end{document}